# IMPROVISED APRIORI ALGORITHM USING FREQUENT PATTERN TREE FOR REAL TIME APPLICATIONS

**Akshita Bhandari[1], Ashutosh Gupta[2], Debasis Das[3]**

[1] *Student, Department of Computer Science and Engineering, NIIT University, Rajasthan, India*

akshita.bhandari@st.niituniversity.in

[2] *Student, Department of Computer Science and Engineering, NIIT University, Rajasthan, India*

ashutosh.gupta@st.niituniversity.in

[3] *Assistant Professor, Department of Computer Science and Engineering, NIIT University, Rajasthan, India*

debasis.das@niituniversity.in

## Abstract

*There are several mining algorithms which have been developed over the years. Apriori Algorithm is one of the most important algorithm which is used to extract frequent itemsets from large database and get the association rule for discovering the knowledge. It basically requires two important things: minimum support and minimum confidence. Firstly, we check whether the items are greater than or equal to the minimum support and we find the frequent itemsets respectively. Secondly, the minimum confidence constraint is used to form association rules. Based on this algorithm, this paper indicates the limitation of the original Apriori algorithm of wasting time and space for scanning the whole database searching on the frequent itemsets, and presents an improvement on Apriori by reducing that wasted time depending on scanning only some transactions by implementing a mathematical formula which initially partitions the set of transactions into clusters and select one particular cluster out of this. Our Algorithm can be used in the library for finding the book that is most frequently read and it can also be used in the grocery shop database by the shopkeeper for finding the itemsets which are frequently sold as this takes lesser time and it's easy to find the items so that shopkeeper can make profit by getting the information of those items which are frequently sold. It gives this result only by using parallel algorithm. The code is implemented in java and the platform used is eclipse. This algorithm's result is generated on Mac using parallel algorithm otherwise it would be similar to the results generated so far by many others. That is how the results are shown and the data structure used in this approach is the frequent pattern tree which can also be used to generate conditional patterns and suitable trees can be drawn for all the items.*

-----------------------------------------------------------------***-----------------------------------------------------------------

## 1. INTRODUCTION

Apriori Algorithm is one of the most popular algorithm in data mining for learning the concept of association rules. It is being used by so many people specifically for transaction operations and also it can be used in real time applications (for instance, grocery shop, general store, library etc.) by collecting the items bought by customers over the time so that frequent itemsets can be generated. Frequent itemsets (itemsets with frequency greater than or equal to a user specified minimum support) can be found very easily because of its combinatorial explosion. Once they are obtained, it is simply easy to generate association rules with confidence greater than or equal to a user specified minimum confidence. [1] With the invent in technology of information and the need for extracting useful information of business people from dataset, [2] data mining and its techniques is appeared to achieve the above goal. As large amount of data is stored in data warehouses, on line analytical process, databases and other repositories of information. If a person tries to search for the information, it can be done manually which may take huge(exponential) amount of time. This is not at all optimum and efficient, so data mining approach is the best way by which this problem can be solved very easily. It is the process in which hidden and kind of interesting patterns are generated from huge amount of data which certainly limits the running time. This data may reach to more than terabytes. In some places Data mining can be termed as knowledge discovery in databases as it generates hidden and interesting patterns, and it also comprises of the amalgamation of methodologies from various disciplines such as statistics, neural networks, database technology, machine learning and information retrieval, etc. Interesting patterns are extracted at reasonable time by using the techniques of knowledge discovery in databases(KDD). Frequent itemsets can be found in various ways: using hash-based technique, partitioning, sampling and using vertical data format since it reduces the running time of the algorithm as it finds the itemsets concurrently. The most outstanding improvement over Apriori would be a method called FP-growh (Frequent-Pattern growth) which succeeded in deleting candidate generation. The architecture of data mining system has the following main components: data warehouse, database or other repositories of information, a server that fetches the relevant data from repositories based on the user's request, [3] knowledge base is used as guide of search according to defined some constraint, data mining engine includes set of essential modules, such as characterization, classification, clustering, association, regression and analysis of evolution. Pattern evaluation module interacts with the modules of data mining for generating interested patterns. Finally, graphical user interfaces (GUI) which allow the user to interact and communicate with the data mining system. This algorithm is basically used to extract useful information from massive amount of data present in repositories and warehouse. For instance, a customer who purchases a pack of bread from the grocery shop also tends to buy the butter from the same shop simultaneously.

## 2. RELATED WORK

Mining of frequent itemsets is an important phase in association mining which discovers frequent itemsets in transactions database. It is essential in many tasks of data mining that try to find interesting patterns from datasets, such as association rules, episodes, classifier, clustering and correlation, etc. [4] Over the time, many algorithms are proposed to find frequent itemsets, but all of them can be catalogued into two classes: candidate generation or pattern growth. Apriori [5] is a representative of the candidate generation approach. It generates the candidate itemsets of length (k+1) based on the frequent itemsets of length (k). The itemset frequency can be defined by counting their occurrences in transactions. Frequent Pattern(FP) - growth, is proposed by Han in 2000, where he stated some useful facts about the FP tree. T.Tassa et al [6] proposed secure mining of Association Rules which is based on the Fast Distributed Mining Algorithm and successfully implementd and developed the techniques and methodologies to solve the problem of distributed association rule mining when items are ordered vertically, [6] he also solved the problem of mining generalized association rules, and the problem of discovering subgroups in the horizontal setting. Mahesh Balaji and G Subrahmanya VRK Rao et al [7] in their paper for IEEE proposed Adaptive Implementation Of Apriori Algorithm for Retail Scenario in Cloud Environment which solves the time consuming problem for retail transactional databases. It aims to reduce the response time significantly by using the approach of mining the frequent itemsets. Yanbin Ye, Chia Chu Chiang et al [8] in their paper proposed A Parallel Apriori Algorithm for Frequent Itemsets Mining. They modified Bodon's implementation and converted it into a parallel approach where the input transactions can be read by a parallel computer. The immediate outcome of a parallel computer on this modified implementation is very well presented. Quiang Yang, Yanhong Hu et al [9] in their paper Applications of Improved Apriori Algorithm on Educational Information identified the main problems in the applications and introduced an improved algorithm. Then this algorithm was used for the data education mining. Ketan Shah, Sunita Mahajan et al [10] in their paper Maximizing the efficiency of Parallel Apriori Algorithm suggested how to maximize the efficiency of the parallel Apriori Algorithm. The paper records and observe the performance of the algorithm over different datasets and over n processors on a commodity cluster of machines. The experiments conducted showed that the parallel algorithm scaled well to the number of processes and also improved the efficiency by effective load balancing. Feng Wang, Yong-Hua Li et al [11] in their paper An Improved Apriori Algorithm Based on the Matrix suggested an improved apriori algorithm based on the matrix. It used the matrix effectively indicating the operations in the database and used the "AND operation" to deal with the matrix to generate the largest frequent itemsets. It doesn't need to scan the database again and again to perform operations and therefore takes less time, and it also reduced the number of candidates of frequent itemsets greatly.

## 3. LIMITATIONS OF APRIORI ALGORITHM

Apriori algorithm suffers from some weaknesses in spite of being clear and simple. The main limitation is costly wasting of time to hold vast number of candidate sets with much frequent itemsets, low minimum support or large itemsets. For instance, if there are $10^4$ frequent 1- itemsets, it may need to generate more than $10^7$ candidates into 2- length which in turn will be tested and accumulated. [4] Furthermore, to detect frequent pattern of size 1000 (e.g.) $V_1, V_2...V_{1000}$, it will have to generate $2^{1000}$ candidate itemsets[1] that yields costly wasting of time in candidate generation as it checks for many more sets from candidate itemsets, also it will scan database many times repeatedly for finding candidate itemsets. Apriori will be very low and inefficient when memory capacity is limited with large number of transactions. The proposed approach in this paper reduces the time spent for searching in the database and performing transactions for frequent itemsets and also reduces the memory space with large number of transactions using partitioning and selecting which is described in detail in the proposed model.

## 4. PROBLEM FORMULATION

Apriori Algorithm takes a lot of memory space and response time since it has exponential complexity eg; if there are 100 transactions then it will have $2^{100}$ itemsets and it also does mining twice. We can somehow reduce the itemsets by frequent itemsets mining (FIM) then it will significantly reduce the time taken but it will take a lot of space, and it will be very inefficient for real time applications eg; if a grocery seller wants to know about the most frequent items purchased or if a person wants to know about the books which are read most frequently in the library, they will have to format their systems again and again as it takes a huge memory space for storing candidate and frequent itemsets. So what can be the solution to minimize it? Also can we minimize the running time of the Algorithm further by using a different approach? How? Explain the approach.

## 5. PROPOSED MODEL

This section will address the improved Apriori ideas, the improved Apriori, an example of the improved Apriori, the analysis and evaluation of the improved Apriori and the experiments. In the process of Apriori, the following definitions are needed:

*6.1 Definition 1* : Suppose $T = \{T_1, T_2, \ldots, T_m\}$, (m, 1) is a set of transactions, $T_i = \{I_1, I_2, \ldots, I_n\}$, (n, 1) is the set of items, and k-itemset = $\{i_1, i_2, \ldots, i_k\}$, (k, 1) is also the set of k items, and k-itemset $\subseteq$ I.

*6.1 Definition 2* : Suppose (itemset) is the support count of itemset or the frequency of occurrence of an itemset in transactions.

*Definition 3* : Suppose $C_k$ is the candidate itemset of size k, and $F_k$ is the frequent itemset of size k.

# 6. ALGORITHM FOR $C_K$ GENERATION

Step1: Scan all transactions to generate $F_1$ table. $F_1$ (items, support, transaction ids)

Step2: Construct by self-join.

Step3: Use $F_1$ to identify the target transactions for .

Step4: Scan the target transactions to generate.

In our proposed approach, we enhance the Apriori algorithm [14] to reduce the time consuming for candidate itemset generation. We first scan all the transactions to generate $F_1$ which contains the items, their support count and Transaction IDs where the items are found. And then we use $F_1$ as a helper to generate $F_2, F_3, \ldots, F_k$. When we want to generate $C_2$, we make a self-join $F_1 * F_1$ to construct 2-itemset C (x, y), where x and y are the items of $C_2$. Before scanning all the transaction records to count the support count of each candidate, use $F_1$ to get the transaction IDs of the minimum support count between x and y, and thus scan for $C_2$ only in these specific transactions. The same thing applies for $C_3$, construct 3-itemset C(x, y, z), where x, y and z are the items of $C_3$ and use $F_1$ to get the transaction IDs of the minimum support count between x, y and z and then scan for $C_3$ only in these specific transactions and repeat these steps until no new frequent itemsets are identified. Now to reduce the memory space when large transactions are there a simple rule can be followed: Let n be the number of nodes in the FP-tree and k be the color of the clusters of the transactions in the database. Now, certainly n > k. If this is the case then k is at most n - 1. Suppose we have 1000 transactions then k will be at most 999. There are so many possibilities of colors and all the colors will be chosen by us. Well, clearly that leads to a bad choice. Now, let n ≥ k as this can also be possible then k will be at most n but still the rule applies as n cannot be less than k because then at each level nodes will have the same color. It must be same if the tree is fully dependent. Since it takes exponential memory space, the possibilities of colors getting generated should be minimized. This can be done by using another mathematical formula for comparing the number of nodes and colors i.e. n > $2^k$. In this case colors will be minimized drastically eg; if n = 1000 now then k will be approximately $\log_2(1000) = 10$. The base 2 signifies that the cluster is getting partitioned into 2 parts and selecting means out of the two only 1 is getting selected. This can be any number of partitions depending on user's choice. User will be having the choice of deciding the base. The value of the base is equal to the number of partitions of the cluster. Using this approach very less memory space is consumed at a time and items can be mines in a lesser amount of time. Hence, it serves the purpose.

# 7. THE IMPROVISED APRIORI ALGORITHM

The improvement of the algorithm can be described as follows:

Step1: //Partition the cluster into groups let this term be n and k be the colors so the loop will be set //for k times. Now, select clusters one at a time.

Step2: //Generate items, their items' support, transaction ids.

Step3: $F_1$ = find_frequent_1_itemsets (T);

Step4: For (k = 2; $F_{k-1} \neq \phi$; k++) {

Step5: //Generate the $C_k$ from the $F_{k-1}$

Step6: $C_k$ = candidates generated from $F_{k-1}$;

Step7: //get the item $I_w$ with minimum support in $C_k$ using $F_1$, (1 ≤ w ≤ k).

Step8: x = Get _ item_ min_ sup ($C_k$, $F_1$);

Step9: // get the target transaction IDs that contain item x.

Step10: Target = get_ Transaction_ ID(x);

Step11: For each transaction t in Target Do

Step12: Increment the count of all items in $C_k$ that are found in Target;

Step13: $F_k$ = items in $C_k$ min_ support;

End; }

# 8. AN EXAMPLE OF THE IMPROVED APRIORI

Assume that a large supermarket tracks sales data by stock-keeping unit (SKU) for each item, such as "butter", "bread", "jam", "coffee", "cheese", "milk" is identified by a numerical SKU. The supermarket has a database of transactions where each transaction is a set of SKUs that were bought together. [2], [12], [13] Let the database of transactions consist of following itemsets:

The transaction set as shown in Table 1. firstly, scan all transactions to get frequent 1-itemset l1 which contains the items and their support count and the transactions ids that contain these items, and then eliminate the candidates that are infrequent or their support are less than the min_ sup as shown in table 2. The frequent 1-itemset is shown in table 3.

The sets which are in bold will be deleted in frequent 2_itemset as shown in table 4. The sets which are in bold will be deleted in frequent 3_itemsets as shown in table 5.

**Table -1:** The Transactions

| Transaction ID | Itemsets |
|---|---|
| T1 | Milk, Cheese |
| T2 | Milk, Coffee, Butter |
| T3 | Jam, Bread |
| T4 | Bread, Butter, Cheese |
| T5 | Coffee, Milk |
| T6 | Milk, Bread, Butter, Jam |
| T7 | Milk, Bread, Butter, Jam, Cheese |

**Table -2:** The candidate 1- itemset.

| Items | Support |
|---|---|
| Milk | 5 |
| Cheese | 3 |
| Coffee | 2 |
| Bread | 4 |
| Butter | 4 |
| Jam | 3 |

**Table -3:** The frequent 1- itemset.

| Items | Support | T_ID |
|---|---|---|
| Milk | 5 | T1,T2,T5,T6,T7 |
| Cheese | 3 | T1,T4,T7 |
| **Coffee** | **2** | **T2,T5** |
| Bread | 4 | T3,T4,T6,T7 |
| Butter | 4 | T2,T4,T6,T7 |
| Jam | 3 | T3,T6,T7 |

**Table -4:** The frequent 2- itemset.

| Items | Support | Min | Found in |
|---|---|---|---|
| **Milk,Cheese** | **2** | **Cheese** | **T1,T7** |
| **Milk, Bread** | **2** | **Bread** | **T6,T7** |
| Milk, Butter | 3 | Butter | T2,T6,T7 |
| **Milk, Jam** | **2** | **Jam** | **T6,T7** |
| Cheese,Bread | 2 | Bread | T4,T7 |
| Cheese,Butter | 2 | Cheese | T4,T7 |
| Cheese, Jam | 1 | Cheese | T7 |
| Bread, Butter | 3 | Bread | T4,T6,T7 |
| Bread, Jam | 3 | Bread | T3,T6,T7 |
| **Butter, Jam** | **2** | **Jam** | **T6,T7** |

**Table -5:** The frequent 3- itemset.

| Items | Support | Min | Found in |
|---|---|---|---|
| **Milk, Butter, Bread** | **2** | **Bread** | **T6,T7** |
| **Milk, Butter, Jam** | **2** | **Jam** | **T6,T7** |
| **Bread, Butter, Jam** | **2** | **Bread** | **T6,T7** |

The next step is to generate candidate 2-itemset from L1 split each itemset in 2-itemset into two elements then use l1 table to determine the transactions where you can find the itemset in, rather than searching for them in all transactions. For example, let's take the first item in table.4 (Milk, Cheese), in the original Apriori we scan all 7 transactions to find the item (Milk, Cheese); but in our proposed improved algorithm we will split the item (Milk, Cheese), into Milk and Cheese and get the minimum support between them using L1 has the smallest minimum support. After that we search for itemset (Milk, Cheese) only in the transactions T1 the minimum confidence, and then generate all candidate association rules.In the previous example, if we count the number of scanned transactions to get (1, 2, 3)-itemset using the original Apriori and our improved Apriori, we will observe the obvious difference between number of scanned transactions with our improved Apriori and the original Apriori. From the table 6, number of transactions in1-itemset is the same in both of sides, and whenever the k of k-itemset increase, the gap between our improved Apriori and the original Apriori increase from view of time consumed, and hence this will reduce the time consumed to generate candidate support count. To get support count for every itemset, here Cheese, and T7. For a given frequent itemset LK, T4, find all non-empty subsets that satisfy the minimum confidence, and then generate all candidate association rules.In the previous example, if we count the number of scanned transactions to get (1, 2, 3)-itemset using the original Apriori and our improved Apriori, we will observe the obvious difference between number of scanned transactions with our improved Apriori and the original Apriori. From the table 6, number of transactions in1-itemset is the same in both of sides, and whenever the k of k-itemset increase, the gap between our improved Apriori and the original Apriori increase from view of time consumed, and hence this will reduce the time consumed to generate candidate support count.

**Graph -1:** The FP-tree of the itemsets

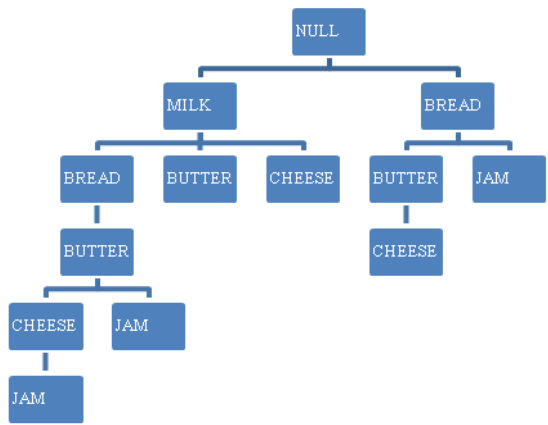

| Nomenclature | | |
|---|---|---|
| M | Milk : 5 | |
| B | Bread : 4 | |
| T | Butter : 4 | |
| C | Cheese : 3 | |
| J | Jam : 3 | |

The final output of the FP-Tree is as shown in Graph:1. And the minimum support count is 3. Now we will find the frequent patterns from the FP-Tree.It's trivial.The items of the database and their frequency of occurrences is shown in Table:2 for each item.First and foremost, we need to prioritize all the itemsets according to their frequency of occurrences and then we will see each item one by one from bottom to top.The items can be listed as:

Then we see Jam. First we need to find the conditional pattern base for Jam:3. If you wonder how 3 comes, it is due to frequency of occurrence of Jam. Now go to Graph 1 and check the Jams. There are 3 Jams and one occurrence for each. Now traverse bottom to top and get the branches which have Jams with the occurrence of Jam. We got 3 branches and they are MBTC: 1, B: 1, MBT: 1.To ensure that you correctly got all the occurrences of Jam in FP-Tree add occurrences of each branch and compare with the occurrences listed above.For Jam you get 1+1+1 = 3 so we got it correct. Then we consider Cheese.And this way we can ensure the correctness for all.Also we can generate conditional pattern bases by rewriting the occurrences for all the branches and finding which item occurs most frequently and then we can delete all other branches except that and only that branch will remain in the FP-Tree which we can draw again for Jam and likewise for all other items.

## 9. THE ANALYSIS AND EVALUATION OF THE IMPROVED APRIORI

Apriori Algorithm used to scan the database twice but this paper presents an improvement on it by using parallel algorithm and the concept of partitioning. It presents a mathematical formula for selecting the cluster as there are many clusters. The code is implemented in java and the platform used is eclipse. The architecture used is mac os x for calculating the running time of the algorithm. The data structure which is introduced in the paper is frequent – pattern tree which is used for finding out the frequent itemsets and also used for generating the conditional patterns. The analysis shows that the time consumed in improved Apriori in each group of transaction is less than the original Apriori, and the difference increases more and more as the number of transactions increases. With the increase in the number of transactions the rate also increases. The average of reducing time rate in the improved Apriori is 67.87%. Apriori is 71.28%. The memory space is reduced by using the partitioning approach which partitions the clusters initially and select one particular cluster out of this. It is an improvement as earlier the algorithm took exponential space but now it is reduced greatly. The mathematical formula for calculating the value of k is shown above.

**Table -6:** The Time Reducing Rate of Improved Apriori according to the number of transactions

| T | Original Apriori(s) | Improved Apriori(s) | Time reducing rate(%) |
|---|---|---|---|
| T1 | 1.776 | 0.654 | 63.17% |
| T2 | 8.221 | 3.982 | 51.56% |
| T3 | 6.871 | 2.302 | 66.49% |
| T4 | 11.940 | 2.446 | 79.51% |
| T5 | 82.558 | 17.639 | 78.63% |

**Table -7:** The Time Reducing Rate of Improved Apriori according to the value of minimum support

| Min_Sup | Original Apriori(s) | Improved Apriori(s) | Time reducing rate (%) |
|---|---|---|---|
| 0.02 | 6.638 | 1.047 | 84.22% |
| 0.04 | 1.855 | 0.398 | 78.54% |
| 0.06 | 1.158 | 0.28 | 75.82% |
| 0.08 | 0.424 | 0.183 | 56.83% |
| 0.10 | 0.382 | 0.149 | 60.99% |

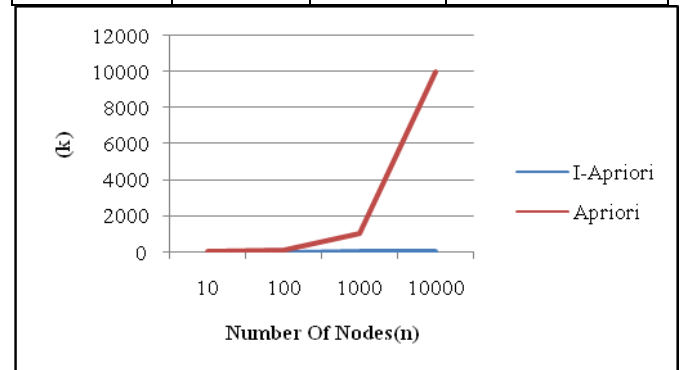

**Fig -1:** Graph between number of nodes and k

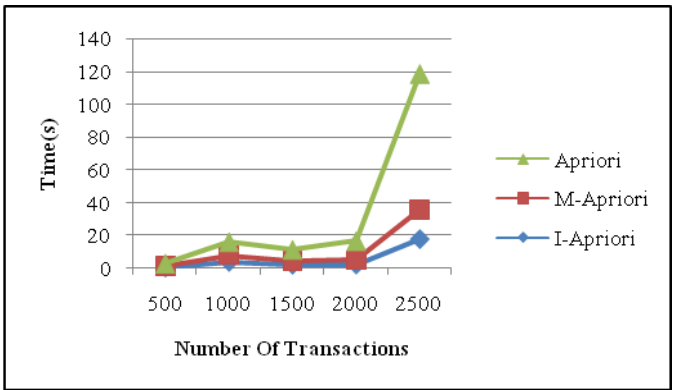

**Fig -2:** Time consuming comparison for different groups of transactions.

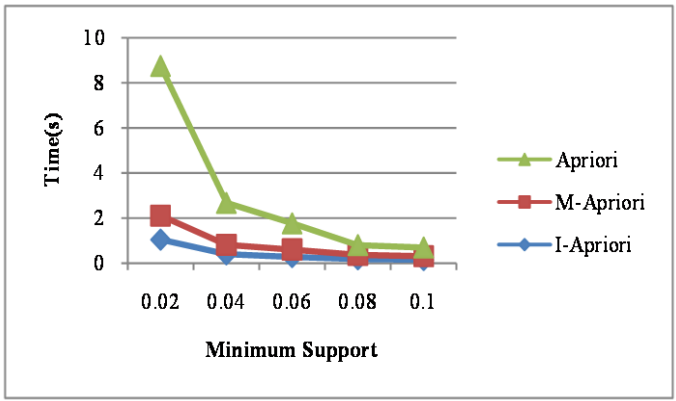

**Fig -3:** Time consuming comparison for different values of minimum support.

We developed an implementation for original Apriori and our improved Apriori, and we collect 5 different groups of transactions as the following:

- T1: 500 transactions.
- T2: 1000 transactions.
- T3: 1500 transactions.
- T4: 2000 transactions.
- T5: 2500 transactions.

The first experiment compares the time consumed of original Apriori, and our improved algorithm by applying the five groups of transactions in the implementation. The result is shown in Figure 2. The second experiment compares the time consumed of original Apriori, and our proposed algorithm by applying the one group of transactions through various values for minimum support in the implementation. The result is shown in Figure 3. As we observe in figure 2, that the time consuming in improved Apriori in each group of transactions is less than it in the original Apriori, and the difference increases more and more as the number of transactions increases. Table 6 shows that the improved Apriori reduce the time consuming by 63.17% from the original Apriori in the first group of transactions T1, and by 78.63% in T5. As the number of transactions increase the rate is increased also. The average of reducing time rate in the improved Apriori is 67.87%. As we observe in figure 3, that the time consuming in improved Apriori in each value of minimum support is less than it in the original Apriori, and the difference increases more and more as the value of minimum support decreases. Table 7 shows that the improved Apriori reduce the time consuming by 84.22% from the original Apriori where the minimum support is 0.02, and by 60.99% in 0.10. As the value of minimum support increase the rate is decreased also. The average of reducing time rate in the improved Apriori is 71.28%.

## 10. CONCLUSIONS

In this paper, the memory space is drastically reduced when large number of transactions are performed from the data warehouses and repositories and an improvised Apriori is proposed by reducing the time consumed in transactions scanning for candidate itemsets and also by reducing the number of transactions to be scanned. Whenever the k of k-itemset increases, the gap between our improved Apriori and the original Apriori increases from view of time consumed, and whenever the value of minimum support increases, the gap between our improved Apriori and the original Apriori decreases from view of time consumed. The time consumed to generate candidate support count in our improved Apriori is less than the time consumed in the original Apriori; our improved Apriori reduces the time consuming by 67.87%. Hence, this approach is far more efficient than the original apriori algorithm as it uses the approach of parallel algorithm and clustering method by which the memory space is reduced and it can be successfully used in the real time applications especially in the library as it can save a lot of time by giving all the information about those books which are frequently read.

## ACKNOWLEDGEMENT

I am really grateful to all the people who directly or indirectly helped me in this research work and also I am thankful to my parents who always helped in with all the ups and downs in my life.

## BIOGRAPHIES

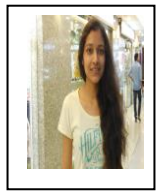

Ms. Akshita Bhandari is a B.Tech III Year student of Computer Science Department in NIIT University. She is the topper of her university. Her area of research is data structures and algorithms. Her research paper has been accepted for ICICT Conference 2014 and Elsevier Journal.

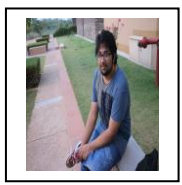

Mr. Ashutosh Gupta is a B.Tech III Year student of Computer Science Department in NIIT University. He is the topper of his university. His area of research is data structures and algorithms and computational biology. His research paper has been accepted for ICICT Conference 2014 and Elsevier Journal.

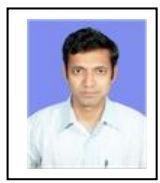

Mr. Debasis Das, is Assistant Professor in Computer Science and Engineering Department in NU. He is pursuing Ph. D in VANET from Department of Computer Science and Engineering, IIT Patna, Bihar, India. He received his M. Tech in Computer Science and Engineering degree from School of Computer Engineering, KIIT University, Bhubaneswar, India in 2010. His research interests include Computer Network, Vehicular Network, Algorithm, Network Security and Cellular Automata. He has published 2 research papers in International journals & a book.